\documentclass[doublecol]{epl2}

\usepackage{graphicx}

\usepackage{color}
\usepackage{amsmath}
\usepackage{amssymb}

\title{Field Dependent Superfluid Density in the Optimally Doped SmFeAsO$_{1-x}$F$_y$ Superconductor}
\shorttitle{Field Dependent Superfluid Density in SmFeAsO$_{1-x}$F$_y$} 

\author{S.~Weyeneth\inst{1} \thanks{E-mail: \email{wstephen@physik.uzh.ch}} \and M.~Bendele\inst{1,2} \and R.~Puzniak\inst{3} \and F.~Mur\'anyi\inst{1} \and A.~Bussmann-Holder\inst{4} \and N.~D.~Zhigadlo\inst{5} \and S.~Katrych\inst{5} \and Z.~Bukowski\inst{5} \and J.~Karpinski\inst{5} \and A.~Shengelaya\inst{6} \and R.~Khasanov\inst{2} \and H.~Keller\inst{1}}
\shortauthor{S.~Weyeneth \etal}

\institute{                    
  \inst{1} Physik-Institut der Universit\"at Z\"urich, Winterthurerstrasse 190, CH-8057 Z\"urich, Switzerland\\
  \inst{2} Laboratory for Muon Spin Spectroscopy, Paul Scherrer Institute, CH-5232 Villigen PSI, Switzerland\\
  \inst{3} Institute of Physics, Polish Academy of Sciences, Aleja Lotnik\'ow 32/46, PL-02-668 Warsaw, Poland\\
  \inst{4} Max-Planck-Institut f\"ur Festk\"orperforschung, Heisenbergstrasse 1, D-70569 Stuttgart, Germany\\
  \inst{5} Laboratory for Solid State Physics, ETH Zurich, Schafmattstrasse 16, CH-8093 Zurich, Switzerland\\
  \inst{6} Department of Physics, Tbilisi State University, Chavchavadze 3, GE-0128 Tbilisi, Georgia}

\abstract{The magnetic field dependence of the in-plane magnetic penetration depth $\lambda_{ab}$ for optimally doped SmFeAsO$_{1-x}$F$_y$ was investigated by combining torque magnetometry, SQUID magnetometry, and muon-spin rotation. The results obtained from these techniques show all a pronounced decrease of the superfluid density $\rho_{\rm s}\propto \lambda_{ab}^{-2}$ as the field is increased to 1.4~T. This behaviour of $\rho_{\rm s}$ is analysed within a two-band model with self-consistently derived coupled gaps and $\rho_{\rm s}=\rho_{\rm s1}+\rho_{\rm s2}$, where $\rho_{\rm s1}$ related to the larger gap is field independent, and $\rho_{\rm s2}$ related to the smaller gap is strongly suppressed with increasing field.}

\begin{document}

\maketitle

\begin{figure*}[t!]
\vspace{-0cm}
\centerline{\includegraphics[width=0.6\linewidth]{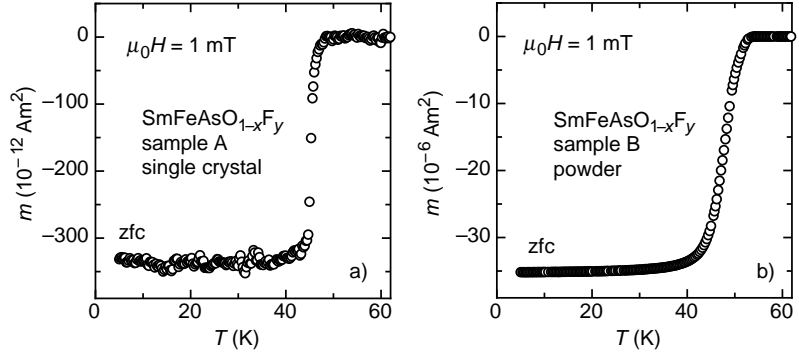}}
\caption{Measured magnetic moment $m$ of SmFeAsO$_{1-x}$F$_y$ as a function of temperature $T$ in 1~mT magnetic field: a) single-crystal sample~A, b) powder sample~B. The rather sharp transition at $T_{\rm c}$ demonstrates the good quality of our samples.}
\label{fig0}
\end{figure*}

After the discovery of superconductivity in LaFeAsO$_{1-x}$F$_x$ \cite{One} with a transition temperature $T_{\rm c}\simeq26$~K, a whole new family of iron-based superconductors was found with a maximum $T_{\rm c}\simeq55$~K for SmFeAsO$_{1-x}$F$_{y}$ \cite{Two}, where $y\leqslant x$ due to a possible oxygen deficiency in the chemical composition \cite{Three}. Various experiments indicate multi-gap superconductivity within the family $RE$FeAsO$_{1-x}$F$_{y}$ ($RE=$ rare-earth element) \cite{Three, Four, Five, Six, Seven, Eight}. Furthermore, the magnetic penetration depth anisotropy, $\gamma_\lambda=\lambda_{c}/\lambda_{ab}$, increases with decreasing temperature, in contrast to the upper critical field anisotropy, $\gamma_H=H_{\rm c2}^{||ab}/H_{\rm c2}^{||c}$, which decreases with decreasing temperature \cite{Nine, Ten, Eleven}, similar to those of the two-gap superconductor MgB$_2$ \cite{Twelve, Thirteen}, although with reversed slopes \cite{Fourteen, Fifteen, Sixteen}. Here $\lambda_{i}$ and $H_{\rm c2}^{||i}$ denote the magnetic penetration depth and the upper critical field components along the crystallographic direction $i$ ($ab$-plane or $c$-axis).\\\indent
Besides of the influence of the magnetic field $H$ and temperature $T$ on the anisotropy, the direct influence of $H$ and $T$ on $\lambda$ is essential in probing multi-gap superconductivity \cite{Seventeen}. However, it is difficult to obtain reliable experimental evidence for multi-gap superconductivity from the temperature dependence of $\lambda$ in samples containing magnetic ions such as SmFeAsO$_{1-x}$F$_y$. Importantly, the superfluid density $\rho_{\rm s}=n_{\rm s}/m^*$, with $n_{\rm s}$ being the superfluid carrier density and $m^*$ the effective carrier mass, can be probed directly by measuring $\lambda^{-2}\propto\rho_{\rm s}$.\\\indent
In the two-gap superconductor MgB$_2$ \cite{Seventeen} $\rho_{\rm s}$ was found to be field dependent due to a suppression of the superfluid density in the band with the smaller gap \cite{Eighteen, Nineteen}. A similar dependence was observed for NbSe$_2$ \cite{Twenty, Twentyone}, V$_3$Si \cite{Twentytwo}, YNi$_2$B$_2$C \cite{Twentythree}, and La$_{1.83}$Sr$_{0.17}$CuO$_4$ \cite{Twentyfour} and again suggested to stem from either multi-gap superconductivity or to be related to changes of the internal field distribution of the vortex lattice.\\\indent
Previous studies of $\lambda_{ab}$ in LaFeAsO$_{1-x}$F$_x$ \cite{Twentyfive} and in SmFeAsO$_{1-x}$ and NdFeAsO$_{1-x}$ \cite{Twentysix} by muon-spin rotation ($\mu$SR) experiments indicated a field dependent $\rho_{\rm s}$ up to 0.6~T in the $RE$FeAsO$_{1-x}$F$_{y}$ family. Also in the related Ba(Fe$_{0.926}$Co$_{0.074}$)$_2$As$_2$ compound a field dependent $\rho_{\rm s}$ was reported for fields up to 0.2~T \cite{Twentyseven}, and being interpreted in terms of multi-gap superconductivity. However, high field investigations of SmFeAsO$_{0.8}$F$_{0.2}$ single crystals between 3~T and 30~T infer a field independent $\rho_{\rm s}$ for SmFeAsO$_{1-x}$F$_y$ \cite{Twentyeight}. This obvious controversy is clarified in this letter by extended investigations of the magnetic field dependence of $\lambda_{ab}$ in the SmFeAsO$_{1-x}$F$_y$ system. As the magnetic penetration depth $\lambda_{ab}$ derived from $\mu$SR data must be carefully evaluated in order to demonstrate that the observed magnetic field dependence of $\rho_{\rm s}$ is intrinsic and not influenced by artificial contributions from the magnetic field distribution, additional experimental methods to investigate $\lambda_{ab}$ are desired. Magnetisation as a global probe is less sensitive to the microscopic details of the magnetic field distribution inside a superconductor which is measured in $\mu$SR experiments. Additionally in magnetic torque experiments $\lambda_{ab}$ is measured in terms of the average torque amplitude and almost insensitive to vortex pinning effects. Thus, a combination of magnetic and $\mu$SR experiments is eligible to provide convincing experimental evidence for the magnetic field dependence of $\lambda_{ab}$. Here, we report on a combined magnetic torque, SQUID magnetisation, and $\mu$SR study of the magnetic penetration depth, evidencing a pronounced magnetic field dependence of $\lambda_{ab}$ in optimally doped SmFeAsO$_{1-x}$F$_y$ up to 1.4 T.\\\indent
In this work single crystal and powder SmFeAsO$_{1-x}$F$_{y}$ samples were used, both of them synthesised by high-pressure techniques \cite{Three, Twentynine}. During the growth process, the composition SmFeAsO$_{1-x}$F$_{y}$ was fine tuned by varying $x$ and $y$, in order to achieve a doping level close to optimal \cite{Three}. The single crystal of nominal composition SmFeAsO$_{0.8}$F$_{0.2}$ (sample A) with an onset of $T_{\rm c}\simeq47$~K had approximate dimensions of $90\times80\times5$~$\mu$m$^3$. The powder sample of nominal composition SmFeAsO$_{0.85}$F$_{0.15}$ (sample B) with an onset of $T_{\rm c}\simeq54$~K was pressed into a pellet of cylindrical shape with 4~mm diameter and 2~mm height. Both samples were characterised by X-ray diffraction with similar properties as reported earlier \cite{Three, Twentynine}. Figure~\ref{fig0} displays the low-field magnetic moment $m$ of sample A and B, measured in a SQUID magnetometer MPMS-XL ({\it Quantum Design}) using the reciprocating sample option (RSO) with $\mu_0H\simeq1$~mT in zero-field cooled (zfc) mode \cite{Three, Twentynine}. Both samples exhibit ideal diamagnetism at low temperature. The transition width (defined by the temperature range between 10\% and 90\% of ideal diamagnetism observed in the zfc low-field $m(T)$ curve) is found to be 2~K and 6~K for the samples A and B, respectively, which demonstrates the good quality of the samples studied here.\\\indent
The single crystal sample A was investigated by torque magnetometry in order to determine $\lambda_{ab}(H)$. All measurements were performed using a custom-made torque magnetometer, previously employed to determine the anisotropic properties of tiny single crystals of iron pnictides \cite{Nine, Ten}. The magnetic torque $\vec{\tau}=\mu_0(\vec{m}\times\vec{H})$, related to the magnetic moment $\vec{m}$, was recorded as a function of the angle $\theta$ between the magnetic field $\vec{H}$ and the $c$-axis of the crystal. During the measurement, $H$ was rotated clock- and counter clockwise around the sample at constant temperature $T$. Due to the present irreversibility, a vortex-shaking technique \cite{Nine, Thirty} was applied with a 200~Hz electromagnetic field of a few mT perpendicular to $H$. The clockwise $(\theta^+)$ and counter clockwise $(\theta^-)$ torque scans were averaged according to $\tau_{\rm rev}(\theta)=[\tau(\theta^+)+\tau(\theta^-)]/2$, in order to further diminish irreversibility effects. A small sinusoidal normal state background of anisotropic paramagnetic origin was subtracted from all data by a symmetrising procedure \cite{Ten, Twentyeight}.\\\indent
\begin{figure}[t!]
\centering
\vspace{-0cm}
\centerline{\includegraphics[width=0.9\linewidth]{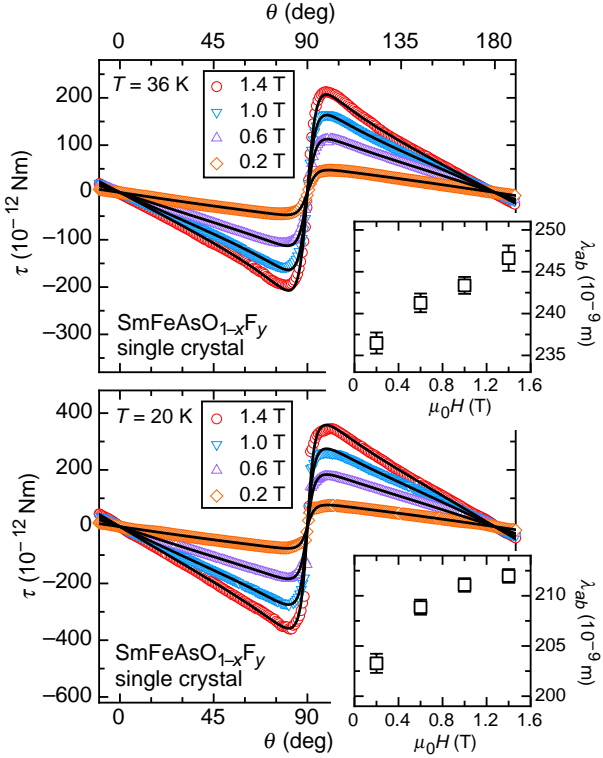}}
\caption{(Colour online) Reversible torque for the single crystal SmFeAsO$_{1-x}$F$_y$ (sample A), measured at 36~K and 20~K in various magnetic fields. Solid lines are calculated using Eq.~(\ref{eq1}), from which $\lambda_{ab}$ was extracted. The inserts show the magnetic field dependence of $\lambda_{ab}$.}
\label{fig1}
\end{figure}
A theoretical expression for the angular dependence of the torque, involving the two anisotropy parameters $\gamma_\lambda$ and $\gamma_H$ independently, was derived within Ginzburg-Landau theory by Kogan \cite{Thirtyone}:
\begin{eqnarray}\label{eq1}
\tau(\theta)&=&-\frac{V\Phi_0H}{16\pi\lambda_{ab}^2}\left(1-\frac{1}{\gamma_\lambda^2}\right)\frac{\sin(2\theta)}{\epsilon_\lambda}\\\nonumber
& &\cdot\Bigg[\ln\left(\frac{\eta H_\mathrm{c2}^{||c}}{H}\frac{4{\rm e}^2\epsilon_\lambda}{(\epsilon_\lambda+\epsilon_H)^2}\right)\\\nonumber
& &-\frac{2\epsilon_\lambda}{\epsilon_\lambda+\epsilon_H}\left(1+\frac{\epsilon'_\lambda}{\epsilon'_H}\right)\Bigg]. \nonumber
\end{eqnarray}
Here $\Phi_0$ is the magnetic flux quantum and $V$ the volume of the crystal. The scaling function $\epsilon_i(\theta)=[\cos^2(\theta)+\gamma_i^{-2}\sin^2(\theta)]^{1/2}$ with $i=\lambda,H$ is different for $\gamma_\lambda$ and $\gamma_H$, and $\epsilon'_i(\theta)$ denotes the derivative with respect to $\theta$. The torque data were analysed with Eq.~(\ref{eq1}). From earlier work \cite{Nine, Ten} the temperature dependence of $H_\mathrm{c2}^{||c}$ was determined, and the anisotropy parameters were found to be field independent up to 1.4~T. Since Eq.~(\ref{eq1}) contains multiple correlated parameters, making a simultaneous analysis of all quantities difficult, all parameters except $\lambda_{ab}$ were kept fixed in the studied field range at the same values reported in \cite{Ten}, thereby reducing the scattering of $\lambda_{ab}$. Because, upper critical fields in optimally doped SmFeAsO$_{1-x}$F$_{y}$ are reported to exceed tens of Tesla \cite{Three, Four, Twentyeight}, this restriction to fix $H_\mathrm{c2}^{||c}$ for different $H$ at the same temperature during the data analysis appears reasonable since the field distribution of the vortex lattice at such low magnetic fields is close to the London limit. Representative experimental torque data collected at two temperatures and various magnetic fields together with the calculated curves [Eq.~(\ref{eq1})] are presented in Fig.~\ref{fig1}. At 20~K and 36~K $\lambda_{ab}$ increases substantially with increasing magnetic field (see inserts to Fig.~\ref{fig1}). For $T<20$~K the presence of disturbing pinning effects \cite{Nine} made the analysis meaningless, while for $T>36$~K the torque signal was too weak to extract reliable values of $\lambda_{ab}$, especially in the relevant low field regime.\\\indent
Additional SQUID magnetisation and $\mu$SR investigations were performed on the larger powder sample (sample B). For anisotropic layered superconductors the effective magnetic penetration depth $\lambda_{\rm eff}$ of unoriented powder samples can be related  to $\lambda_{ab}$ via $\lambda_{\rm eff}\simeq1.31\lambda_{ab}$ \cite{Thirtytwo}. For optimally doped SmFeAsO$_{1-x}$F$_y$ we find values of $\gamma_\lambda$, exceeding 7 at $T_{\rm c}$ and increasing to 19 at zero temperature \cite{Nine, Ten}, thus confirming this approximation  and being in accordance with other experiments which reveal comparable values for $\gamma_\lambda$ in related compounds \cite{Twentyeight, Thirtythree}. Field dependent SQUID measurements of the magnetic moment at 45 K are shown in the upper panel of Fig.~\ref{fig2}. The analysis of these data is based on the average of the magnetisation for increasing and decreasing magnetic field with $\lambda_{ab}^{-2}$ being derived as \cite{Seventeen}
\begin{equation}
\lambda_{ab}^{-2}\propto\frac{\partial m}{\partial\ln(H)}=H\frac{\partial m}{\partial H}.\label{eq2}
\end{equation}
From Eq.~(\ref{eq2}) $\lambda_{ab}^{-2}$ was determined and is shown in the lower panel of Fig.~\ref{fig2}. Obviously, $\lambda_{ab}^{-2}$ at 45~K decreases continuously as the field is increased up to 1.4~T, in good agreement with the torque result (see Fig.~\ref{fig3}).\\\indent
\begin{figure}[t!]
\centering
\vspace{-0cm}
\centerline{\includegraphics[width=0.9\linewidth]{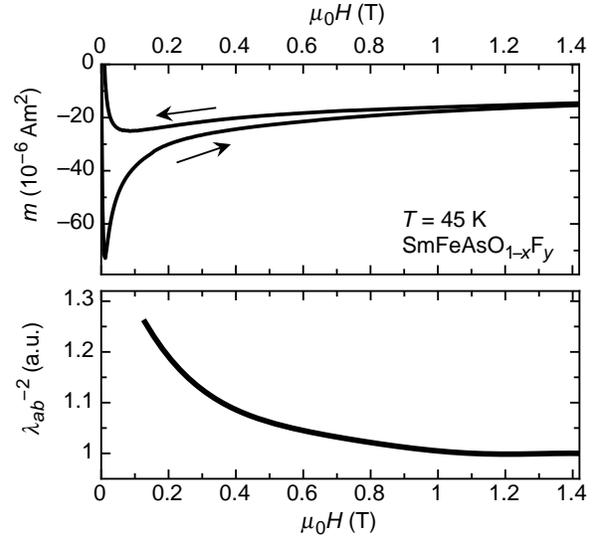}}
\caption{Results of the SQUID investigation for sample B (powder) of SmFeAsO$_{1-x}$F$_y$. Upper panel: Field dependence of the magnetic moment $m$ at 45~K measured for increasing and decreasing magnetic field. Lower panel: Field dependence of $\lambda_{ab}^{-2}$ derived by Eq.~(\ref{eq2}).}
\label{fig2}
\end{figure}
In order to substantiate our main result on the magnetic field dependence of $\lambda_{ab}$ obtained  on single crystals, the powder sample B was further studied by transverse-field (TF) and longitudinal-field (LF) $\mu$SR at the Paul Scherrer Institute (PSI) (For detailed information on the TF and LF $\mu$SR techniques see {\it e.g.} \cite{Thirtyfour}). Previous $\mu$SR investigations of $RE$FeAsO$_{1-x}$F$_y$ \cite{Thirtyfive, Twentysix, Thirtysix, Thirtyseven} indicate that: (i) The muons stop at two lattice sites, one is close to the magnetic moments of the Fe ions in the FeAs layers, whereas the other is near the spacing layers containing the rare-earth ions \cite{Thirtyfive, Twentysix, Thirtysix}. (ii) The magnetic moments of the Sm ions order antiferromagnetically at temperatures below $\sim5$~K, leading to a pronounced local-field broadening of the magnetic field distribution \cite{Twentysix, Thirtysix}. In addition, for temperatures above 5~K the finite correlation of the magnetic moments leads to thermally activated spin-fluctuations and causes a damping of the TF time spectra \cite{Twentysix, Thirtyseven}. (iii) Consistent with the above two points, the $\mu$SR time spectra of SmFeAsO$_{1-x}$F$_y$ consist of a slow and a fast relaxing component \cite{Thirtyfive, Thirtysix}. Combining these three points and invoking, according to calculations by Hayano {\it et al.} \cite{Thirtyeight}, the magnetic correlation of the Sm moments to be similar in LF and TF experiments, the total transverse-field asymmetry $A(t)$ of the time evolution of the muon-spin polarisation in SmFeAsO$_{1-x}$F$_y$ can be written as \cite{Twentysix}
\begin{equation}
A(t)=\left(A_{\rm s}{\rm e}^{-\Lambda_{\rm s}t}+A_{\rm f}{\rm e}^{-\Lambda_{\rm f}t}\right){\rm e}^{-\sigma^2t^2/2}\cos(g_\mu Bt+\phi).
\label{eq3}
\end{equation}
Here $A_{\rm s}$ ($A_{\rm f}$) and $\Lambda_{\rm s}$ ($\Lambda_{\rm f}$) are the asymmetry and the depolarisation rate of the slow (fast) relaxing component, respectively, $B$ is the magnetic induction at the muon sites, $g_\mu$ is the muon gyromagnetic ratio, $\phi$ is the initial phase of the muon-spin ensemble, and $\sigma$ is the Gaussian measure of the second moment of the magnetic field distribution within the sample.\\\indent
All $\mu$SR time spectra were fitted simultaneously by Eq.~(\ref{eq3}) with the ratio $A_{\rm s}/A_{\rm f}$ as a common parameter and $\Lambda_{\rm s}$ and $\Lambda_{\rm f}$ as inidividual parameters for each data set. The total asymmetry $A_{\rm s}+A_{\rm f}$ was kept constant within each set of the data. The temperature dependent depolarisation rates $\Lambda_{\rm s}$ and $\Lambda_{\rm f}$ in Eq.~(\ref{eq3}) are both assumed to have the form \cite{Twentysix}
\begin{equation}
\frac{1}{\Lambda_i(T)}=\frac{1}{\Lambda_i(0)}+C\exp(-E_0/k_{\rm B}T), \quad\quad i={\rm s,f},
\label{eq4}
\end{equation}
with $E_0$ being the activation energy for both relaxations and $C$ a constant. Our analysis yields $\Lambda_{\rm s}(0)\simeq0.20(5)$~$\mu$s$^{-1}$, $\Lambda_{\rm f}(0)\simeq2.71(7)$~$\mu$s$^{-1}$, and $E_0\simeq20(1)$~meV. This procedure includes the background contribution due to the correlation of the Sm magnetic moments. Since the Sm moments order antiferromagnetically below $\sim5$~K \cite{Twentysix, Thirtysix}, an analysis for $T\leqslant10$ K  was omitted in this work. From Eq.~(\ref{eq3}) the two-component Gaussian relaxation rate $\sigma$ was obtained: $\sigma=(\sigma^2_{\rm sc}+\sigma^2_{\rm nm})^{1/2}$, where $\sigma_{\rm nm}$ is the temperature independent nuclear moment contribution, and  $\sigma_{\rm sc}$ is the superconducting relaxation rate due to the muon-spin depolarisation in the vortex lattice \cite{Twentysix}. The values of $\lambda_{ab}\simeq\lambda_{\rm eff}/1.31$ were derived from $\sigma_{\rm sc}$ according to \cite{Thirtynine}
\begin{eqnarray}
\sigma_{\rm sc} [\mu {\rm s}^{-1}]&=&4.83\times10^4(1-H/H_{\rm c2})\\
&{\rm }&\cdot\left[1+1.21\left(1-\sqrt{H/H_{\rm c2}}\right)^3\right]\lambda_{\rm eff}^{-2}[{\rm nm}],\nonumber
\label{eq5}
\end{eqnarray}
with the same values of $H_{\rm c2}$ as used for the torque experiment. The magnetic field dependence of $\lambda_{ab}^{-2}$, derived from the $\mu$SR data together with the results from extended torque studies, are presented in Fig.~\ref{fig3}. Evidently, $\lambda_{ab}^{-2}$ is substantially suppressed with increasing magnetic field at all temperatures studied and follows, when normalised to $\lambda_{ab}^{-2}(1.4~{\rm T}/\mu_0)$, on a universal curve (see Fig.~\ref{fig4}).\\\indent
\begin{figure}[t!]
\centering
\vspace{-0cm}
\centerline{\includegraphics[width=0.9\linewidth]{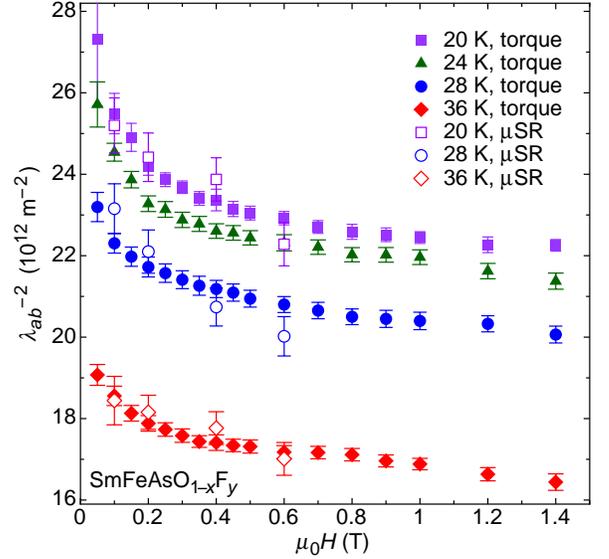}}
\caption{(Colour online) Field dependence of $\lambda_{ab}^{-2}\propto \rho_{\rm s}$ of SmFeAsO$_{1-x}$F$_y$ derived from the magnetic torque (sample A, closed symbols) and $\mu$SR (sample B, open symbols) experiments.}
\label{fig3}
\end{figure}
The present results as obtained from all three techniques demonstrate very clearly a decrease of $\lambda_{ab}^{-2}$ with increasing magnetic field. From Figs.~\ref{fig2}, \ref{fig3}, and \ref{fig4} we estimate a suppression of the superfluid density by 20\% when the field increases from 0~T up to 1.4~T. Assuming a single-gap scenario, for $H\ll H^{||c}_{\rm c2}$, only a nodal superconductor is expected to show a field dependent superfluid density \cite{Forty, Twentyfive, Thirtyseven, Twentysix}. Since however SmFeAsO$_{1-x}$F$_y$ is nodeless \cite{Eight, Fortyone}, alternative explanations need to be considered. An influence of $H_{\rm c1}$ on the determination of $\lambda_{ab}^{-2}$ can be excluded, since for an estimated value of $\lambda_{ab}(0)\simeq200$~nm the corresponding $H^{||c}_{\rm c1}(0)\propto \lambda_{ab}^{-2}(0)$ is of the order of $0.05$~T, well below the lowest field studied here. A systematic error in the determination of $\lambda_{ab}$ caused by pinning appears to be unlikely as well: (i) In SQUID magnetometry pinning contributes strongly at low temperatures. The field dependence of the magnetic moment was performed at 45~K, where pinning effects are considerably reduced. (ii) In the $\mu$SR experiments all data were obtained in field cooled vortex-lattice states. From the field dependence of the magnetic field distribution in the mixed state, a distortion of the $\mu$SR spectra caused by pinning can be excluded \cite{Twentysix}. (iii) In the torque experiment, a shaking technique was applied in order to minimise irreversibility effects \cite{Thirty}. Thus, the determination of $\lambda_{ab}$ is not affected by pinning. Furthermore, the derived $\lambda_{ab}$ is mainly susceptible to the average torque amplitude which is almost insensitive to pinning, in contrast to $H^{||c}_{\rm c2}$, which was fixed during the data analysis \cite{Nine, Ten}.\\\indent
\begin{figure}[t!]
\centering
\vspace{-0cm}
\centerline{\includegraphics[width=0.9\linewidth]{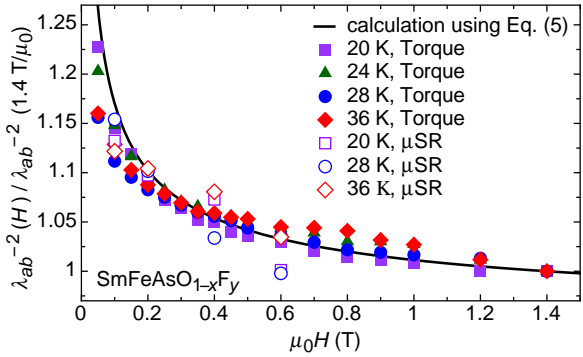}}
\caption{(Colour online) Universal behaviour of $\rho_{\rm s}$ observed for both samples (A and B) of SmFeAsO$_{1-x}$F$_y$ for all investigated temperatures, normalised to its value at the maximum field of 1.4~T. The black line is calculated using Eq.~(\ref{eq6}). The meaning of the symbols is the same as in Fig.~\ref{fig3}}
\label{fig4}
\end{figure}
As already mentioned above, the observed magnetic field dependence of $\lambda_{ab}^{-2}$ is absent in nodeless single-gap superconductors \cite{Fortytwo} but has been observed in various multi-band superconductors, where the cases of MgB$_2$ \cite{Seventeen, Eighteen, Nineteen} and La$_{1.83}$Sr$_{0.17}$CuO$_4$ \cite{Twentyfour} have been analysed in detail. The analogy to these systems suggests that related physics apply here as well. Assuming that SmFeAsO$_{1-x}$F$_y$ is a fully-gapped superconductor \cite{Eight, Fortyone}, a two-gap model \cite{Fortythree} is used to calculate the superfluid density $\rho_{\rm s}=\rho_{\rm s1}+\rho_{\rm s2}$, stemming from the coupled gaps ($\Delta_1,\Delta_2$).  In this approach, the zero-temperature gaps are calculated self-consistently in relation to the total superfluid density $\rho_{\rm s}$. The best agreement is found by fixing the values $\Delta_1(0)=13.83$~meV and $\Delta_2(0)=5.26$~meV, close to the s-wave gap values as obtained by point-contact spectroscopy \cite{Six, Seven}, performed on samples from the same source \cite{Twentynine}. With increasing magnetic field the superfluid density is mainly suppressed in the band with the small gap ($\rho_{\rm s2},\Delta_2$) where the corresponding intra-band interaction approaches zero with increasing field. Note however, that the gap itself is not zero due to finite inter-band interactions analogous to MgB$_2$ \cite{Eighteen, Nineteen}. The total superfluid density $\rho_{\rm s}(H)$ is expressed as a sum of a contribution from the band with the large gap, $\rho_{\rm s1}$, which is independent of the field, and a field dependent contribution from the band with the small gap, $\rho_{\rm s2}\propto1/\sqrt{H}$. The field independence of $\rho_{\rm s1}$ stems from the same physics as in MgB$_2$ \cite{Eighteen, Nineteen} and the $1/\sqrt{H}$ dependence of $\rho_{\rm s2}(H)$ is based on the multi-gap analysis of La$_{1.83}$Sr$_{0.17}$CuO$_4$, where the coupled gaps have different field dependencies, namely, the $d$-wave gap being proportional to $\sqrt{H}$ \cite{Fortyfour} and the $s$-wave gap having a $1/\sqrt{H}$ dependence \cite{Fortyfive}. Since the $s$-wave symmetry of the SmFeAsO$_{1-x}$F$_y$ superconductor is well justified \cite{Eight}, the $1/\sqrt{H}$ dependence is plausible. Thus, the expression $\rho_{\rm s}(H)/\rho_{\rm s1}$ is derived as
\begin{equation}
\rho_{\rm s}(H)/\rho_{\rm s1}=1+C(H/H_0)^{-1/2}/\rho_{\rm s1},
\label{eq6}
\end{equation} 	
where $C$ denotes the amplitude of $\rho_{\rm s2}(H)$ and $H_0$ is the characteristic field for suppressing $\rho_{\rm s2}$. A fit to the normalised superfluid density data by Eq.~(\ref{eq6}) leads to an estimation of $(C/\rho_{\rm s1})^2\mu_0H_0\simeq6.1$~mT. The calculated result is shown in Fig.~\ref{fig4}, in comparison with the normalised data, with obviously good agreement between both.\\\indent
It is important to emphasise that although $\rho_{\rm s}$ is suppressed by $20\%$ in small fields, $T_{\rm c}$ is almost not changing on the same field scale, consistent with high values of $H_{\rm c2}$. This can be understood in terms of a suppression of the intra-band coupling within the band of the small gap, having a pronounced influence on the corresponding $\rho_{\rm s}$, but almost no effect on $T_{\rm c}$ \cite{Eighteen, Nineteen}.\\\indent
In conclusion, we observe a magnetic field dependence of the superfluid density in the iron-pnictide superconductor SmFeAsO$_{1-x}$F$_y$. This experimental observation can be consistently described within a coupled two-gap model, where the superfluid density related to the large gap is field independent, whereas the one of the small gap is strongly suppressed with increasing magnetic field. Together with the temperature dependence of the anisotropy parameters, multi-gap superconductivity is well reflected in the thermodynamic behaviour of SmFeAsO$_{1-x}$F$_y$.\\\indent

\acknowledgments
We thank S.~Str\"assle for his help in preparing the manuscript and B.~M.~Wojek for helpful discussions and useful comments. This work was in part performed at the S$\mu$S at Paul Scherrer Institute (Switzerland) and was partially supported by the Swiss National Science Foundation, the NCCR program MaNEP, the Polish Ministry of Science and Higher Education, within the research project for the years 2007-2010 (No. N N202 4132 33), and the Georgian National Science Foundation grant GNSF/ST08/4-416.

\end{document}